# Pulsed-laser epitaxy of metallic delafossite PdCrO$_2$ films

Jong Mok Ok[1], Matthew Brahlek[1], Woo Seok Choi[2], Kevin M. Roccapriore[3], Matthew F. Chisholm[3], Soyeun Kim[4], Changhee Sohn[1,4], Elizabeth Skoropata[1], Sangmoon Yoon[1], Jun Sung Kim[5,6], Ho Nyung Lee[1]*

[1]Materials Science and Technology Division, Oak Ridge National Laboratory, Oak Ridge, TN 37831, U.S.A.

[2]Department of Physics, Sungkyunkwan University, Suwon 16419, Korea

[3]Center for Nanophase Materials Sciences, Oak Ridge National Laboratory, Oak Ridge, TN 37831, U.S.A.

[4]Department of Physics, Ulsan National Institute of Science and Technology, Ulsan 44919, Republic of Korea

[5]Department of Physics, Pohang University of Science and Technology, Pohang, Republic of Korea

[6]Center for Artificial Low Dimensional Electronic Systems, Institute for Basic Science (IBS), Pohang, Republic of Korea.

*Correspondence should be addressed to hnlee@ornl.gov

**Abstract:** Alternate stacking of the highly conducting metallic layers with the magnetic triangular layers found in the delafossite PdCrO$_2$ is an excellent platform for discovering intriguing correlated quantum phenomena. Thin film growth of this material may enable not only tuning the basic physical properties beyond what bulk materials can exhibit, but also developing novel hybrid materials by enabling interfacing dissimilar materials, yet this has proven to be extremely challenging. Here, we report the epitaxial growth of metallic delafossite PdCrO$_2$ films by pulsed laser epitaxy (PLE). The fundamental role the growth conditions, epitaxial strain, and chemical and structural characteristics of the substrate are investigated by growing under various growth conditions and on various types of substrates. While strain plays a large role in improving the crystallinity, the direct growth of epitaxial PdCrO$_2$ films without impurity phases was not successful. We attribute this difficulty to both the chemical and structural dissimilarities between the substrates and volatile nature of PdO layer, which inhibit nucleation of the delafossite phase. This difficulty was overcome by growing CuCrO$_2$ buffer layers before PdCrO$_2$ were grown. Unlike PdCrO$_2$, CuCrO$_2$ films were grown across a wide range of conditions. It was found that a single monolayer of CuCrO$_2$ was sufficient to grow the correct PdCrO$_2$ phase. This result indicates that the epitaxy of Pd-based delafossites is extremely sensitive to the chemistry and structure of the interface, necessitating near perfect substrate materials. The resulting films are commensurately strained, highly metallic, and show antiferromagnetism that emerges at 37 K and persists down to as thin as 3.6 nm in thickness. This work provides key insights into advancing the epitaxial growth of the broader class of metallic delafossites for both studying the basic physical properties and new technologies.

$AB$O$_2$ delafossites are a class of materials offering a wide range of physical properties due to the combination of electronic conduction and magnetism [1, 2]. Unique to this class of materials is the layered structure consisting of alternating $A$ and $B$O$_2$ layers as shown in Fig. 1(a). The natural heterostructuring of the $A$ and $B$O$_2$ layers makes the delafossite a key material system to study the interplay among itinerate behaviors of electrons with localized non-collinear magnetism. Further, the intrinsically long mean free path, leading to an extremely high conductivity as well as the two-dimensional (2D) nature of the transport, make this material system for transformative applications [3, 4]. The $A$ layer is composed of closed-packed cations with $A$ = Cu, Pd, Ag, or Pt, whereas the $B$O$_2$ layer is composed of slightly distorted edge-sharing $B$O$_6$ octahedra with $B$ = Al, Cr, Fe, Co, or Rh [1, 2, 5]. The delafossites with $A$ = Cu, Ag are semiconductors and have long been studied as $p$-type transparent conducting oxides [6, 7]. This characteristic is contrasted by delafossites with $A$ = Pd or Pt, which are highly metallic and exhibit conductivities as high as copper [3, 4, 8]. Further, the triangular in-plane connectivity gives rise to non-collinear antiferromagnetic states for materials, such as CuFeO$_2$ [9], CuCrO$_2$ [10], AgNiO$_2$ [11], and PdCrO$_2$ [12]. While several studies were reported from bulk materials, including a large magnetoresistance [13], unconventional anomalous Hall effect [14, 15], field-induced magnetic transition [9, 11], and magnetoelectric effect [16, 17], many interesting questions, including the strain tuning, proximity effect, and dimensional control of interlayer coupling, are yet to be understood as there have been only few attempts in epitaxial growth of metallic delafossites [18, 19, 20, 21]. Among the delafossites, PdCrO$_2$ is highly attractive owing to strong coupling between highly conducting 2D Pd layers and non-collinear antiferromagnetic CrO$_2$ layers, and, consequently, exhibits unusual transport properties [14, 15, 22, 23]. However, the growth of Pd-based delafossites thin films have been shown to be challenging due to the volatile nature of PdO as well as challenges controlling nucleation of with impurity phase associated with reduction of the Pd [19, 20].

Here, we have systematically studied the epitaxial synthesis of the delafossite PdCrO$_2$ by pulsed laser epitaxy (PLE) on various substrates. By systematically tuning the growth conditions, growth temperature ($T$), oxygen partial pressure ($P_{O2}$), and laser fluence ($J$), it was found that the formation of high-quality, phase-pure PdCrO$_2$ epitaxial films requires a delicate balance of the growth conditions. Although the conditions that favor the growth of $PdCrO_2$ are nominally independent of the substrate, the highest quality $PdCrO_2$ was found to grow on low-lattice-mismatched delafossite buffer layers, which significantly reduced the appearance of impurity phases. The quality of the buffer layer was found to be the limiting factor for the growth of $PdCrO_2$. This indicates that a new generation of bulk crystals will ultimately enable synthesis of high quality $PdCrO_2$ epitaxial thin films.

The pulsed-laser epitaxy of $PdCrO_2$ films used a sintered polycrystalline target. Based on previous reports on bulk synthesis [2, 24], the polycrystalline $PdCrO_2$ target was prepared using a combination of Pd, $PdCl_2$ and $LiCrO_2$. These powders were mixed in a stoichiometric ratio then sintered at 900 °C in a vacuum furnace, which resulted in $PdCrO_2$ mixed with LiCl. The mixture was then ground and washed in distilled water for 30 minutes to remove the LiCl byproduct. Note that LiCl is highly soluble in water, unlike $PdCrO_2$. The resulting $PdCrO_2$ polycrystalline powder was then dried by heating to 120 °C in air for approximately 12 hours. This power was then pelletized and sintered at 800 °C in atmosphere to form the final target. The single crystal was used to compare the optical properties with the thin films. For the $PdCrO_2$ single crystal, a mixture of the obtained polycrystalline $PdCrO_2$ and NaCl flux were annealed at 900 °C for 24 hours and slowly cooled to 800 °C as described elsewhere [15, 24]. For the films, the growth conditions were varied ($T = 500 - 800$ °C, $P_{O2} = 10 - 1000$ mTorr, and $J = 1 - 2$ J/cm$^2$), whereas the repetition rate of the KrF excimer laser ($\lambda = 248$ nm) was fixed at 10 Hz. After the growth, the samples were cooled to room temperature in $P_{O2} = 100$ Torr. $CuCrO_2$ thin films, which were used as a buffer layer, were grown directly on $Al_2O_3$ at $T = 700$ °C under an oxygen pressure of $P_{O2} = 10$ mTorr. It is noted that the growth of the $CuCrO_2$ films was achieved over a wide growth window. The crystal structure was characterized by X-ray diffraction (XRD) using a four-circle high-resolution x-ray diffractometer (X'Pert Pro, PANalytical; Cu $K\alpha_1$ radiation), and the thickness of the film ($d$) was calibrated using X-ray reflectivity (XRR). The surface morphology measurements were made with atomic force microscopy (Veeco Dimension 3100). High-angle annular dark-field scanning transmission electron microscopy (HAADF-STEM) images were collected with a 5[th] order aberration-corrected NION UltraSTEM 200 operated at 200kV using a 30 mrad convergence semi-angle. The transport properties were measured using the van der Pauw configuration with a 14 T Quantum Design PPMS using aluminum wires directly wire-bonded to PdCrO$_2$ films. Optical properties of PdCrO$_2$ thin films, PdCrO$_2$ single crystal, and SrTiO$_3$ substrate were measured using spectroscopic ellipsometry for UV-visible range (M-2000, J. A. Woollam Co.) with photon energies between 0.74 and 5.86 eV at an incident angle of 65°. For both the films and the substrate infrared ellipsometry (IR-VASE, J.A. Woollam) was used for energy below 0.74 eV. The spectra were fitted using a two-layer model (film/substrate) to obtain physically reasonable dielectric functions. For single crystal, the complex optical conductivity was extracted using Kramers-Kronig analysis after measuring near-normal reflectance (Vertex 80v FT-IR spectrometer, Bruker).

PdCrO$_2$ has a rhombohedral structure (space group R-3m) with lattice parameters of $a = b = 2.930$ Å, and $c = 18.087$ Å ($\alpha = \beta = 90°$, $\gamma = 120°$) [1, 12]. The triangular in-plane geometry requires the use of substrates with a triangular (hexagonal) symmetry. This surface symmetry can be obtained from a variety of triangular structures, for example, (111) oriented cubic perovskites that are available with a wide range of lattice constants. As shown in Fig. 1(b), (111)-SrTiO$_3$, (0001)-Al$_2$O$_3$, (111)-MgO, and (0001)-4H-SiC were chosen. Figure 1(b) shows the lattice parameters of various delafossites and commercially available single crystals as potential buffer layers or substrates. It is noted that there have been reports on successful synthesis of bulk materials, including CuCrO$_2$, CuAlO$_2$, CuFeO$_2$, CuGaO$_2$, PdCoO$_2$, PdRhO$_2$ and PtCoO$_2$ [1, 2]. Among them, a few Cu-based delafossites, including CuAlO$_2$ [6] and CuCrO$_2$ [25], were successfully grown as thin films. The lattice mismatches ($\delta(\%) = (a_s - a_f)/a_s \times 100$, where $a_s$ is the lattice parameter of the substrate) of PdCrO$_2$ ($a_{PCO} = 2.930$ Å) with various substrates are indicted in parentheses. Note that the surface lattice size ($a_S$) of (111)-oriented cubic substrates is taken as triangular unit to mate with the oxygen lattice of the delafossite, as illustrated in Fig. 1(b). Overall, this approach gives a wide range of lattice mismatches, i.e., (111)-SrTiO$_3$ ($\delta = -6.1\%$), (0001)-Al$_2$O$_3$ ($\delta = -5.9\%$), (111)-MgO ($\delta = 1.7\%$), and 4H-SiC ($\delta = 4.8\%$). The majority of previous studies of thin-film growth of delafossites have focused on Al$_2$O$_3$ (0001) as a substrate due to its close lattice match and the hexagonal symmetry as well as the chemical similarity of the oxygen terminated surface of Al$_2$O$_3$ to the $B$O$_2$ terminating surface of the delafossites.

Thus, this selection of substrates allows the influence of strain in achieving high-quality delafossite films to be studied.

To understand the role of the substrate in nucleation of both the delafossite and impurity phases, PdCrO$_2$ films were grown on the substrates mentioned earlier, as well as on an additional buffer layer i.e., single-monolayer PLE-grown CuCrO$_2$ ($\delta$ = 1.3%) (~0.6 nm in thickness) on (0001) Al$_2$O$_3$. Initial growth parameters were optimized for PdCrO$_2$ films grown on Al$_2$O$_3$, i.e. $P_{O2}$ = 100 mTorr, $T$ = 700 °C, and $J$ = 1.5 J/cm$^2$. XRD $2\theta$–$\theta$ scans for ~30 nm thick PdCrO$_2$ films grown under the same optimum condition on all of the substrates and buffer layers are shown in Fig. 2(a). As one can find, the appearance of the 0006 PdCrO$_2$ peak indicates confirms the delafossite phase. The PdCrO$_2$ peaks are the most intense for the substrates and buffer layers with the closest lattice match, Al$_2$O$_3$ and delafossite CuCrO$_2$ buffer layer. The PdCrO$_2$ peaks are suppressed in general, as the lattice mismatch increase. Despite a small mismatch, no sign of PdCrO$_2$ film peaks was observed on MgO substrates. The full width at half maximum (FWHM) from XRD rocking curve $\omega$-scans measured for the 0006 peak is plotted as a function of lattice mismatch in Fig. 2(b). This result shows similar dependence on the lattice mismatch. For films grown with a low mismatch, the rocking curve width is low, ~0.1° for Al$_2$O$_3$ and CuCrO$_2$, whereas the FWHM of >1° are found for 4H-SiC and SrTiO$_3$. It is also noted that, while the sign is opposite, PdCrO$_2$/SiC ($\delta$ = 4.8 %) has a similar size of lattice mismatch with PdCrO$_2$/Al$_2$O$_3$ ($\delta$ = –5.9%), but the FWHM of the PdCrO$_2$ phases show a huge difference. This may be because the tensile strain increases oxygen deficiency due to the reduced formation energy or increased oxygen exchange kinetics [26, 27, 28].

The role the substrate plays in nucleation can be additionally seen by comparing the relative intensity of secondary phases. Specifically, in Fig. 2(c) the intensity of the Cr$_2$O$_3$ peak at 39.7° is plotted as a function of the lattice mismatch. The films grown on a buffer layer of CuCrO$_2$ and on 4H-SiC showed highly suppressed Cr$_2$O$_3$ impurity phases, whereas films grown on Al$_2$O$_3$, SrTiO$_3$, and MgO showed Cr$_2$O$_3$ peaks with a much larger intensity. This result shows that the nucleation of Cr$_2$O$_3$ is does not solely depend on the lattice mismatch, but rather is highly dependent on the crystal structure of the substrate. Specifically,

Cr$_2$O$_3$ shares the same crystal structure as Al$_2$O$_3$ and thus seems very energetically favorable to form epitaxially on Al$_2$O$_3$, explaining the dominant formation of Cr$_2$O$_3$. This result further contrasts with films grown on both buffer layer and 4H-SiC, where there is only a very weak Cr$_2$O$_3$ peak, likely indicating that the formation of Cr$_2$O$_3$ phase is unfavorable on these surfaces. In addition to the crystallographic symmetry, it is noted that the formation of Cr$_2$O$_3$ impurities might be suppressed by preventing the loss of volatile PdO by using, for instance, Pd-rich targets as demonstrated for PdCoO$_2$ [18].

In general, for heteroepitaxy, the closer the film and the substrate crystal structures to one another, the higher the film quality. For example, PdCoO$_2$ films grown on Al$_2$O$_3$ have shown epitaxial twins due to equivalent ways to crystallographically connect the delafossite structure to the corundum structure [18, 19, 20]. As the data in Fig. 2 show, the CuCrO$_2$ delafossite buffer layer gives rise to the highest crystalline quality PdCrO$_2$. In addition to the small lattice mismatch and identical crystal symmetry of R-3m, CuCrO$_2$ is electrically insulating, which readily allows transport measurements of PdCrO$_2$.

Thin film of CuCrO$_2$ was grown using PLE on Al$_2$O$_3$ at $T = 700$ °C, $P_{O2} = 10$ mTorr, and $J = 1.5$ J/cm$^2$. An XRD scan is shown in Fig. 3(a) for CuCrO$_2$ where the delafossite 0003$n$ peaks are the only peaks resolved. Moreover, the rocking curve FWHM of the 0006 peak from a thick CuCrO$_2$ film (10 nm) is ~0.1°, indicating a higher film quality than those in previous reports [29]. Using a single monolayer CuCrO$_2$ buffer layer, the growth conditions of PdCrO$_2$ were further mapped systematically. Figure 3(b) shows a full XRD scan and an $\omega$-rocking curve for a PdCrO$_2$ film grown under optimum condition, which confirms a good epitaxy of PdCrO$_2$.

Figure 3(c) summarizes the results for PdCrO$_2$ films grown at different $T$ and $P_{O2}$, where the contour plot indicates rocking curve FWHM values of the 0006 PdCrO$_2$ peak, and the symbols indicate whether the sample is metallic (solid green circles) or insulating (blue stars). The conditions, where there were no PdCrO$_2$ peaks observed, are marked with red crossed circles. From these data, the optimum growth regime is found to be $600 \leq T \leq 700$ °C and $200 \leq P_{O2} \leq 300$ mTorr. Within this optimal growth window, the rocking curve width is minimal, and the films are metallic, whereas the films grown outside of the optimal window are found to be insulating (beyond the instrumental limit) even in case with a good crystallinity.

The insulating nature was found to be due to islands formation which were not fully connected. The rocking curve width ($\Delta\omega = 0.1°$) is nominally the same as the underlying CuCrO$_2$, which indicates that the structural quality is largely limited by the underlying buffer layer. Finally, it is important to point out that these growth conditions are nominally identical to those found for PdCrO$_2$ directly grown on Al$_2$O$_3$, which confirms that formation of the impurities is driven by the character of the surface. In addition, as shown in Fig. 3(d), XRD reciprocal space mapping confirmed that this film is fully strained. HAADF-STEM image shown in Fig. 3(e) for PdCrO$_2$ grown on a single-unit-cell thick CuCrO$_2$ buffered Al$_2$O$_3$ substrate further confirms the high crystallinity of the epitaxial thin film. From this image taken along the <−1100> direction, the layered delafossite crystalline structure with alternatively stacked Pd (bright) and Cr (darker) layers is clearly resolved on the Al$_2$O$_3$ substrate, which is the darker region on the bottom of the image. The atomic structure at the interface is resolvable, for which there are several interesting aspects: First, the interfacial layer that forms on the Al$_2$O$_3$ substrate is found to be the Cr-O layer, followed by the Pd layer, which likely highlights that the Cr-O shares a similar metal-oxygen bond network as Al$_2$O$_3$. Second, noting that the STEM brightness is proportional to the atomic number, $\propto Z^2$, the CuCrO$_2$ buffer layer should be appear at the interfaces with Al$_2$O$_3$ as a dark layer since as Cu (atomic number $Z = 29$) is much lighter than Pd ($Z = 46$). However, there is no obvious sign of the Cu layer from our $Z$-contrast imaging. This result might be due to the fact that first top Cu layer was replaced or mixed with Pd instead of nucleation of a Cr-O layer on top of the Cu layer. Third, the interfacial layer of the substrate shows disorder that might originate from intermixing with the film with the delafossite epilayers. The interfacial structure and role of the nucleation layer can be better understood by more extensive studies utilizing element-specific atomic resolution electron energy loss spectroscopy (EELS), which is left as a future study. Overall, STEM measurements show that the structure is free of impurity phases and of high crystalline quality.

Figure 4 shows thickness dependent transport data for the PdCrO$_2$ films grown on CuCrO$_2$ buffered Al$_2$O$_3$. As shown in Fig. 4(a), our films ($d$ = 3.6–33 nm in thickness) show clear metallic behavior where the resistivity decreases with decreasing temperature, except for the thinnest ($d$ = 3.6 nm) film, of which $\rho(T)$ shows a slight upturn at low temperature, consistent with the onset of a localization transition. Below 3.6 nm, PdCrO$_2$ thin films were no longer conducting due to the finite thickness, typical for conducting oxides such as SrRuO$_3$ [30, 31]. $\rho(T)$ for PdCrO$_2$ films was nearly independent of thickness above ~4 nm, but the value was about an order of magnitude higher ($\rho$ = 100 μΩ·cm at 300 K) than that of a single crystal ($\rho_{single}$ = 8.2 μΩ·cm). The overall quality of the films is captured by the residual resistivity ratio (RRR, the ratio of the room temperature $\rho$ to the low temperature $\rho$). The maximum RRR value is 2.1 for the 33-nm-thick film. This value is comparable to other PdCoO$_2$ thin films grown by pulsed laser deposition [18], but significantly smaller than that from bulk crystals (RRR ~ 200) [15, 24]. This comparison indicates that the film quality may have room for further improvement. $\rho(T)$ also gives insight into the magnetic properties of PdCrO$_2$ films. Bulk PdCrO$_2$ exhibits non-collinear antiferromagnetism with a Neel temperature of $T_N$ = 37.5 K [12, 24], which is reflected in transport measurements as a kink in $\rho(T)$ [15, 24]. This feature is more clearly seen in $d\rho(T)/dT$ shown in Fig. 4(b). We assign the maximum value of $d\rho/dT$ to be $T_N$ and have plotted as a function of thickness as shown in Fig. 4(c). These values are nominally thickness-independent over 3.6–33 nm at $T_N \approx$ 37 K within the estimated error of around ± 5 K, which agrees well with the bulk value.

Finally, optical conductivity spectra ($\sigma_1(\omega)$) of PdCrO$_2$ thin films and a single crystal at the room temperature are shown in Fig. 5(a) and (b), which were extracted from spectroscopic ellipsometry. Overall, $\sigma_1(\omega)$ of the thin films show a consistent behavior with that of the single crystal. First, all films and single crystal exhibited a low-energy Drude peak below around 1 eV, which reflects the metallic nature. There is also several interband transitions observed at 1.3 eV (referred as $E_1$), 2.3 eV ($E_2$), 4.5 eV ($E_3$), and above 6.0 eV ($E_4$). From electronic structure calculations [32], a schematic density of states (DOS) is provided as shown in Fig. 5(c). From the schematic, the $E_1$ peak at 1.3 eV can be attributed to the on-site $d$-$d$ transition between the hybridized Cr orbital states, and the $E_2$ peak at 2.3 eV to the $d$-$d$ transition between the hybridized Cr or Pd orbital states. The $E_3$ and $E_4$ peaks at 4.5 eV and above 6 eV are attributed to the charge transfer transition from the O to the Cr or Pd states. By considering the Drude and Lorentz oscillators, $\sigma_1(\omega)$ fit and the result are shown in Fig. 5(b). The Drude parameters obtained from the fit were plasma frequency, $\omega_p \approx$ 28000 cm$^{-1}$ and scattering rate, $\gamma_i \approx$ 1450±50 cm$^{-1}$ (wavenumber). It is noted that $\omega_p$ of our thin film is in good agreement with that of our PdCrO$_2$ single crystal ($\omega_p \approx 35000$ cm$^{-1}$, $\gamma_i \approx 220$ cm$^{-1}$) and a sister compound PdCoO$_2$ single crystal ($\omega_p \approx 33000$ cm$^{-1}$, $\gamma_i \approx 97$ cm$^{-1}$) [33], but the value of $\gamma_i$ is 7-15 times larger than that from the single crystal. The Drude parameters obtained from the fit are summarized in Table 1. The difference observed in $\gamma_i$ is consistent with the higher resistivity of thin films. The Drude parameters were converted into relaxation time of $\tau \approx 3.65 \times 10^{-15}$ s, carrier density of $n_{3D}^{opt} \approx 1.3 \times 10^{22}$ cm$^{-3}$, mobility of $\mu^{opt} \approx 5.27 \pm 0.5$ cm$^2$/V·s, and DC conductivity of $\sigma^{opt} \approx 9000$ $\Omega^{-1}$·cm$^{-1}$, corresponding to $\rho^{opt} \approx 110$ µΩ·cm. This was consistent with the values from transport measurements. The data were found to be nominally thickness independent, which indicates that the broad electronic properties are preserved down to the thinnest sample. The Drude parameters, however, show clear thickness dependence, as listed in Table 1. The scattering rate $\gamma_i$ was larger for small film thickness. This finding implies that disorder may play a dominant role in the ultrathin limit. It is also noted that the optical conductivity, $\sigma_1(0)$, and DC conductivity, $\sigma_{DC}$, show significant difference for the 3.6 nm-thick sample. This difference is attributed to potential inhomogeneity, e.g., twin boundaries and Cr$_2$O$_3$ impurity phases in the vicinity of the interface.

In summary, we have shown that, albeit on the border of instability, PdCrO$_2$ can be grown epitaxially by pulsed laser epitaxy. Epitaxial strain and the chemical and structural characteristics of the substrate are of key importance to the phase purity of PdCrO$_2$. However, substrates that are at low strain states yet closely match impurity phases strongly enhance formation of these secondary phases. As such, achieving epitaxial PdCrO$_2$ films required a different approach. It is found that the use of a CuCrO$_2$ buffer layer as thin as a single monolayer not only helped reduce the impurity phases, but also improved the crystalline quality of the films in comparison to films grown directly on Al$_2$O$_3$. PdCrO$_2$ epitaxial thin films exhibited a clear magnetic transition down to a thickness of 3.6 nm. Overall, these results show that overcoming significant growth challenges for delafossite family is the first step toward a new generation of complex oxide thin films and creation of novel quantum heterostructures.

## Acknowledgements

This work was supported by the U.S. Department of Energy, Office of Science, Basic Energy Sciences, Materials Sciences and Engineering Division. The transport measurement and analysis were supported by the Computational Materials Sciences Program, and the optical measurement and analysis by W.S.C. was supported by Basic Science Research Program through the National Research Foundation of Korea (NRF-2019R1A2B5B02004546).

## Additional information

Supplementary information is available in the online version of the paper. Reprints and permissions information is available online at ***.

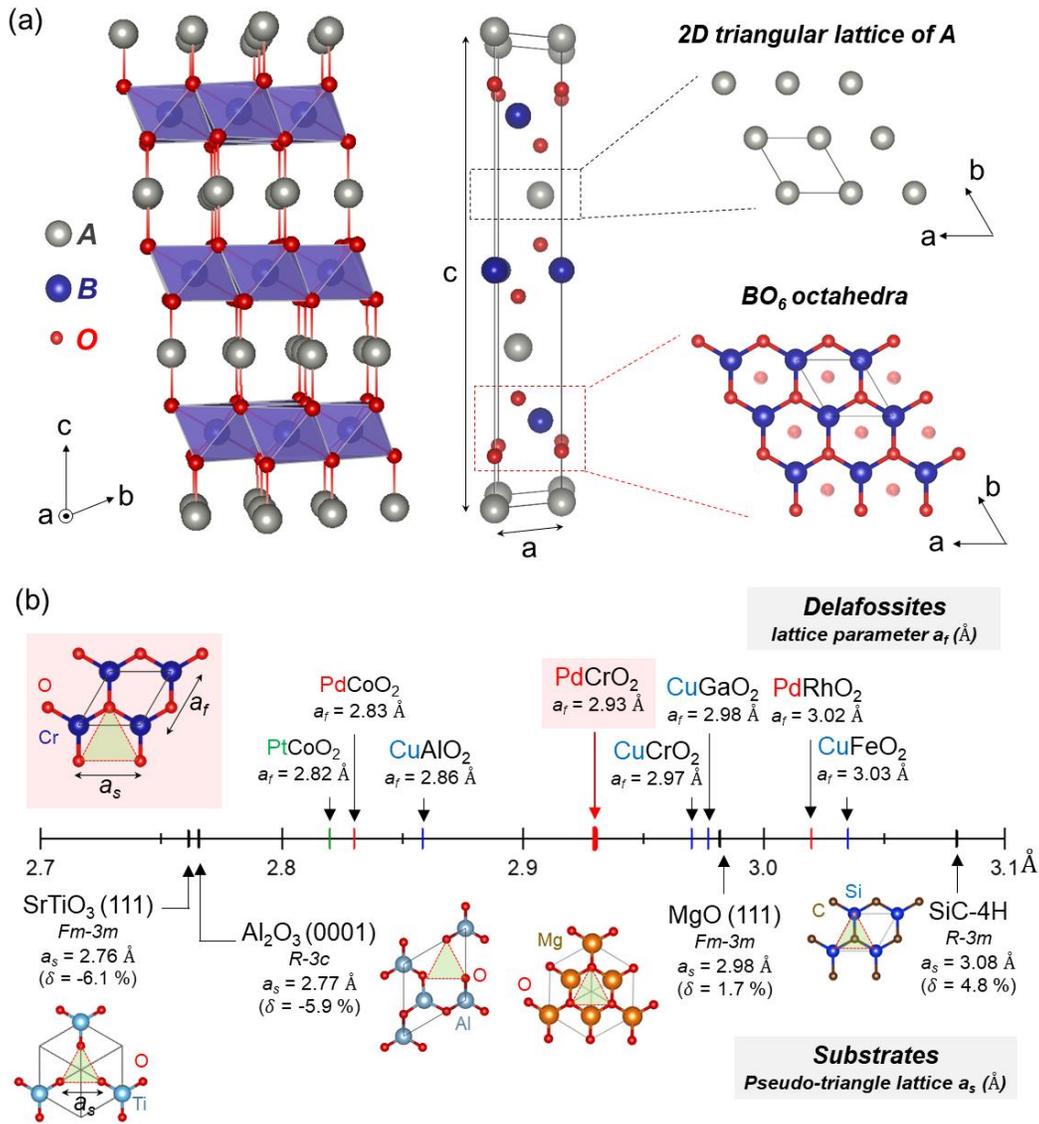

**FIG. 1** (a) Crystal structure of the delafossite $AB$O$_2$ (space group R-3m). $A$, $B$, O atoms are colored silver, blue, and red, respectively. The stacked layers of closed-packed $A$ and edge-sharing $B$O$_6$ octahedra are also shown. (b) Schematic of $B$O$_6$ octahedral surface of the delafossite structure and lattice parameters (top-left corner). Number line showing the lattice parameters of substrates ($a_S$) and delafossite buffer layer materials ($a_f$). Note that the surface lattice sizes ($a_S$) of (111)-oriented cubic substrates are taken as the triangular unit that mate to the oxygen sublattice of the delafossite, as highlighted as a triangular crystal for each substrate.

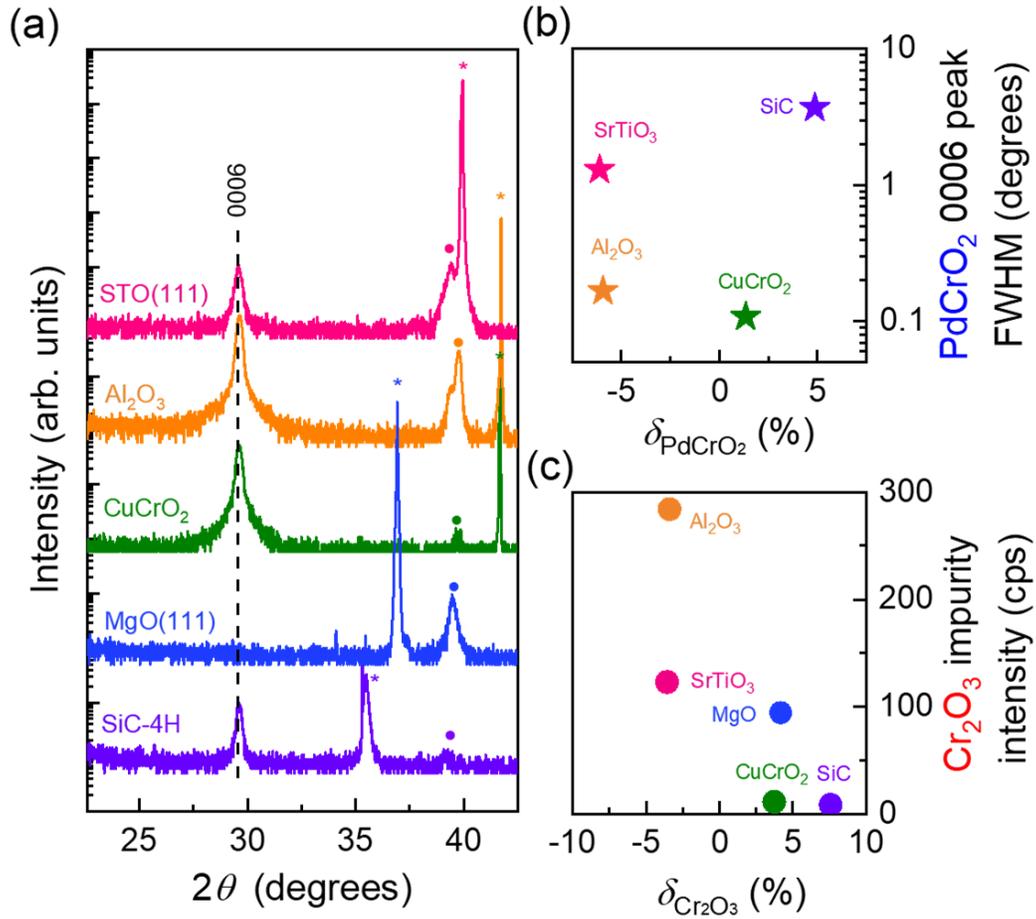

**FIG. 2** (a) XRD scans of PdCrO$_2$ thin films grown on different substrates and buffer layers with the relevant peaks are labeled. The peaks from PdCrO$_2$ thin films, substrates, and Cr$_2$O$_3$ impurity are marked with a dotted line, asterisks (*) and solid dots (•), respectively. (b) Crystallinity of films represented as the full width half maximum (FWHM) of the PdCrO$_2$ 0006 peak as a function of the lattice mismatch of PdCrO$_2$, $\delta_{PdCrO2}(\%) = (a_s - a_{PdCrO2})/a_s \times 100$. (c) XRD intensity of the Cr$_2$O$_3$ impurity peak as a function of the lattice mismatch of Cr$_2$O$_3$, $\delta_{Cr2O3}(\%) = (a_s - a_{Cr2O3})/a_s \times 100$.

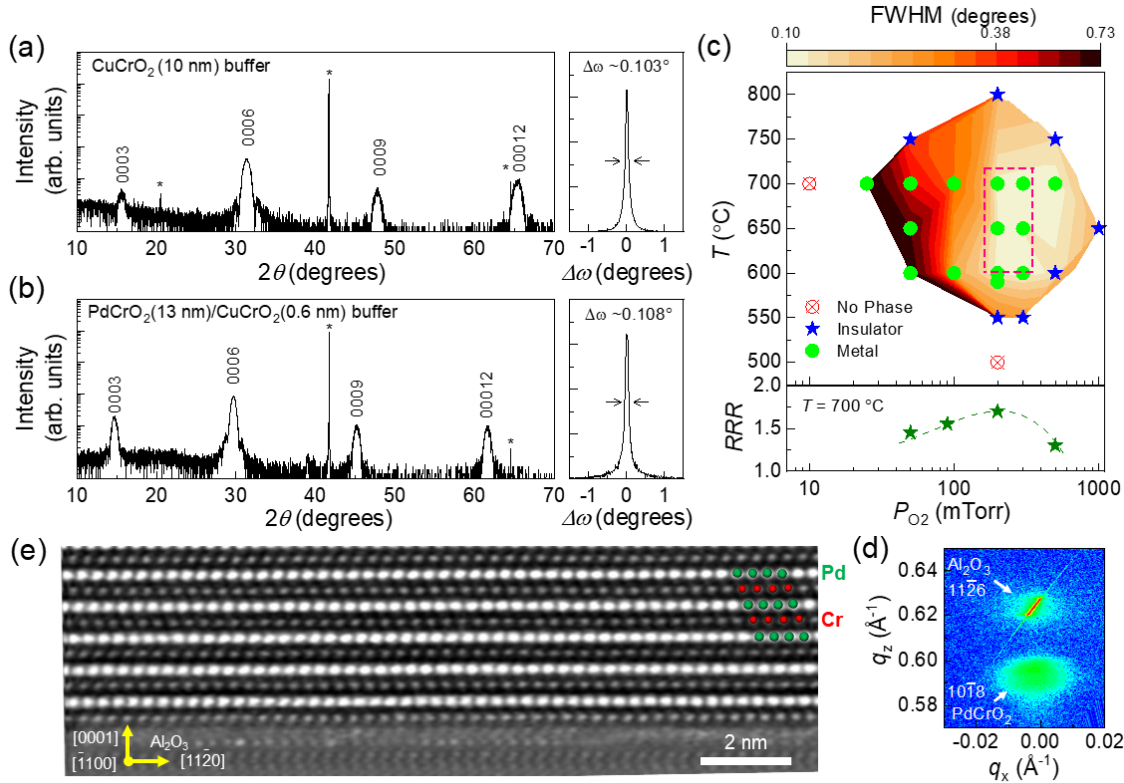

**FIG. 3** (a) XRD 2$\theta$-$\theta$ scans of a CuCrO$_2$ thin film (thickness 10 nm) grown on an Al$_2$O$_3$ (0001) substrate under optimum growth conditions (left). The individual 0003$n$ delafossite peaks are indicated, while the substrate is indicated by an asterisk (*). Corresponding rocking curve for the CuCrO$_2$ 0006 peak is shown on the right. (b) XRD 2$\theta$-$\theta$ scans of a PdCrO$_2$ thin film (thickness 13 nm) grown on a CuCrO$_2$ buffer layer (left). Corresponding rocking curve for the PdCrO$_2$ 0006 peak is shown on the right. (c) Contour plot of FWHM of the PdCrO$_2$ 0006 peak as functions of oxygen partial pressures, $P_{O2}$, and growth temperature, $T$, showing the optimal growth window (top) and oxygen partial pressure dependence of the residual resistivity ratio (RRR) (bottom). (d) XRD reciprocal space map of a 23.5-nm-thick film recorded around the Al$_2$O$_3$ 11-26 reflection and the PdCrO$_2$ 10-18 reflection. (e) A cross-sectional HAADF-STEM image of a PdCrO$_2$ thin film grown on a CuCrO$_2$ buffer layer on Al$_2$O$_3$.

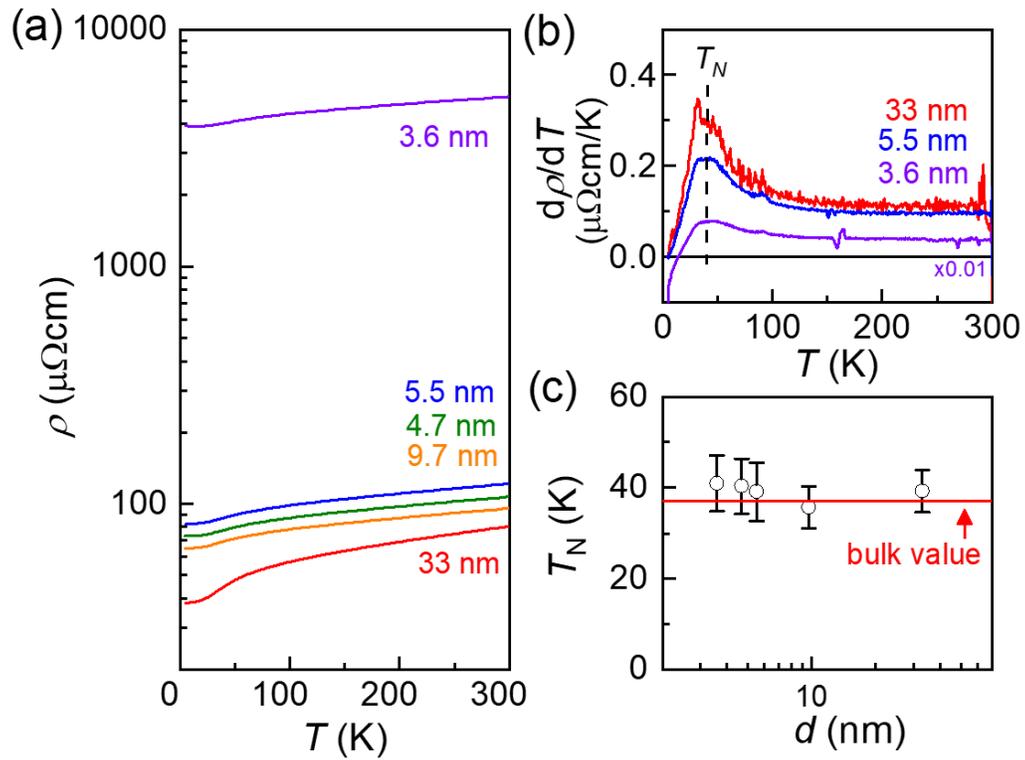

**FIG. 4** (a) Temperature dependence of resistivity of PdCrO$_2$ thin films with various thicknesses of 3.6-33 nm. (b) Temperature dependent d$\rho$/d$T$ data, which show clear kinks at $T_N$ = ~37 K. (c) The antiferromagnetic ordering temperature ($T_N$) from PdCrO$_2$ films with different film thickness, which is compared to the value in bulk shown as the solid red line.

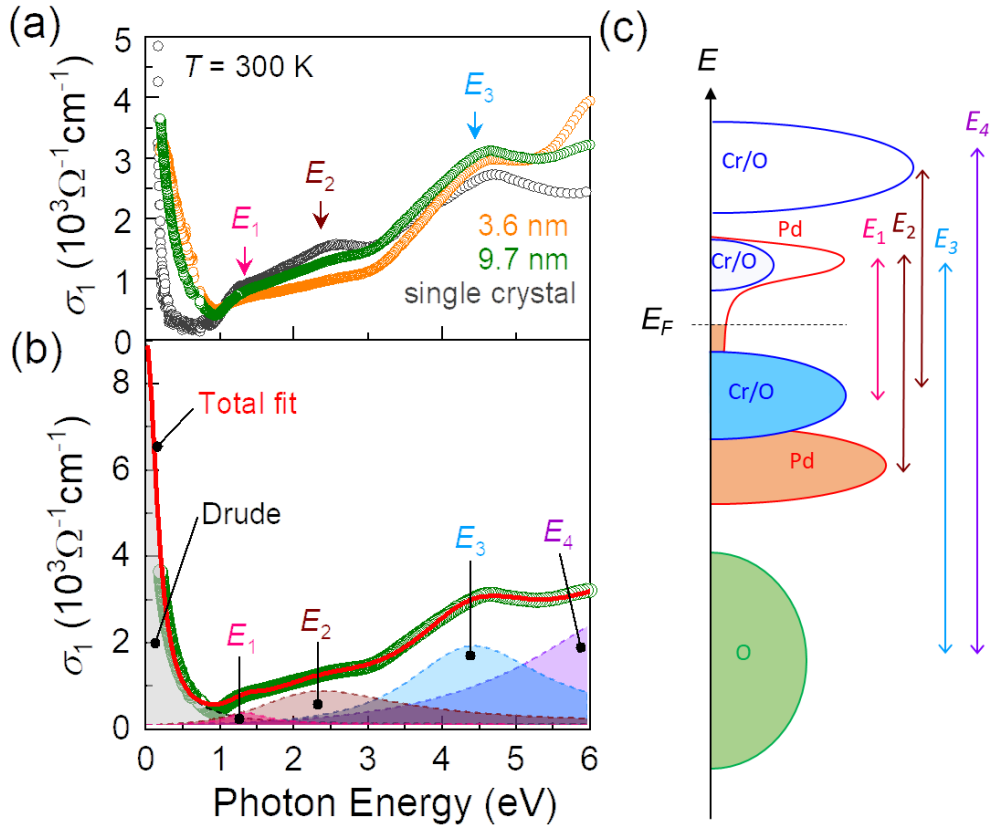

**FIG. 5**. (a) Thickness dependent optical conductivity of PdCrO$_2$ thin films and single crystal, and (b) the corresponding fit is composed of a Drude term (grey colored area) and several Lorentzian features ($E_1$, $E_2$, $E_3$, $E_4$; colored areas). (c) Schematic DOS for in PdCrO$_2$ based on reported density functional theory calculations in Ref. [32]. The expected optical transitions are denoted as $E_1$, $E_2$, $E_3$ and $E_4$.

|  |  | $\omega_p$ [cm$^{-1}$] | $\gamma_i$ [cm$^{-1}$] | $\sigma_1(0)$ [$\Omega^{-1}\cdot$cm$^{-1}$] | $\sigma_{DC}$ [$\Omega^{-1}\cdot$cm$^{-1}$] |
|---|---|---|---|---|---|
| PdCrO$_2$ thin film | 3.6 nm | 25,000 | 3,310 | $4.5\times10^3$ | $0.2\times10^3$ |
|  | 5.5 nm | 28,000 | 2,260 | $6\times10^3$ | $8.3\times10^3$ |
|  | 9.7 nm | 28,000 | 1,450 | $9\times10^3$ | $9.1\times10^3$ |
| PdCrO$_2$ single crystal |  | 35,000 | 220 | $0.8\times10^5$ | $1.22\times10^5$ |
| PdCoO$_2$ single crystal [33] |  | 33,000 | 97 | $1.85\times10^5$ | $3.84\times10^5$ |

**Table 1**. The Drude model parameters of PdCrO$_2$ thin films fitted to the optical conductivity and DC conductivity, which are compared with these values from a PdCrO$_2$ single crystal. Literature values for a PdCoO$_2$ single crystal is also shown as a comparison.